\documentclass[aps,prl,onecolumn,superscriptaddress,groupedaddress]{revtex4}

\usepackage{dcolumn}   
\usepackage{bm}        
\usepackage{amssymb}   
\usepackage{slashed}
\usepackage{graphicx}				
\usepackage{amsmath}
\usepackage{tikz}
\usepackage{tikz-cd}
\usetikzlibrary{matrix,arrows,backgrounds,decorations.pathmorphing}
\usepackage[all]{xy}
\newcommand{\bra}[1]{\ensuremath{\left\langle#1\right|}}
\newcommand{\ket}[1]{\ensuremath{\left|#1\right\rangle}}

\begin{document}
\nocite{*}

\title{Black Holes, Information and the Universal Coefficient Theorem}
\author{Andrei T. Patrascu}
\address{University College London, Department of Physics and Astronomy, London, WC1E 6BT, UK}
\begin{abstract}
General relativity is based on the diffeomorphism covariant formulation of the laws of physics while quantum mechanics is based on the principle of unitary evolution. In this article I provide a possible answer to the black hole information paradox by means of homological algebra and pairings generated by the universal coefficient theorem. The unitarity of processes involving black holes is restored by demanding invariance of the laws of physics to the change of coefficient structures in cohomology. 
\end{abstract}

\maketitle
\section{1. Introduction}
The prescriptions of general relativity and quantum mechanics are taking away most of the absoluteness associated to choices of coordinates, trajectories followed by particles and states of physical systems in the absence of any accessible information about them. The mathematical language of differential form is perfectly suited for dealing with such prescriptions. It is my observation that there still remains an epistemological defect associated to these ideas. Not to all arbitrary conventions has been taken their absolute status away. In fact the connectivity of space is probably the last convention that still is considered absolute by many physicists. Mathematically however, one cannot assign an absolute topology to spacetime in the absence of a method for detecting such a topology. This obstruction is at the origin of several paradoxes and inconsistencies, notably the "information paradox" for black holes.
 Because of this, in order to construct a consistent formulation of physics in a general context, it appears to be necessary for the laws of nature to be specified in a topology-covariant way. In the same way in which the language of differential forms allows us to transfer geometry-independent properties from one manifold to another, the topology-covariant language will allow us to transfer topology-independent properties from one topological space to another. In the same way in which differential forms allow us to construct objects that are coordinate independent, the universal coefficient theorems (and other theorems of homological algebra) will allow us to construct objects that are by construction, independent of choices of coefficients in (co)homology. This will provide us with a tool capable of identifying dualities in a more systematic way.

 In a more practical tone, one of the problems arising in the discussion of black holes in a quantum field theoretical context is the fact that the quantum prescription of unitarity may be lost in processes involving the thermal radiation of black holes [1]. In fact it can be shown that in a semi-classical approximation, each process involving the presence of a horizon may lead to outgoing thermal radiation [2]. An in-falling pure quantum state is then mapped into the external radiation which presents a thermal spectrum thus violating unitarity. I analyze here the origin of this problem and find that the semi-classical approximation is insufficient for a correct quantum description of phenomena involving space-time horizons. In fact, the solution appears to be related to topological properties of the transformation groups considered as acting on the given space. These properties are being analyzed in terms of group (co)homology within the group approach to quantization.

 The discussion continues for a group approach to the second quantization giving rise to a group theoretical analogue for quantum field theory. In order to realize a coefficient independent construction, I use the group extensions in the universal coefficient theorem to derive the connection between homology and cohomology. The resulting pairing between group homology and cohomology will provide us with a coefficient covariant construction as well as additional multipliers that will ensure this at the level of computations.

 
The departure from a strict thermal spectrum in the context of a group approach to quantization for conformally invariant quantum field theories has been discussed in [33]. As the group approach to quantization implies the embedding of the (potentially curved) space $Q$ within a larger differentiable structure containing the phase space of the theory, namely a group $G$ which will from now on drive the quantization procedure, the geometry of the physical space and its physical symmetries as well as possible gauge symmetries will be naturally taken care of in terms of group theoretical effects [34]. 


Also, a given phase space may possibly be embedded in different groups, each of these groups giving rise to different quantum theories having, in particular, non-equivalent vacua. A central extension of this group, namely $G\rightarrow \tilde{G}$ will give rise to the quantizing group, the fundamental structure in the group approach to quantization. The centrally extended group $\tilde{G}$ plays a key role in characterizing the vacuum states in the curved space $Q$. In general, standard quantum field theories in curved spacetime do not have a preferred definition of particles. The origins of this ambiguity lie in the infinite, arbitrary, non-equivalent directions of polarization for the infinite dimensional Heisenberg-Weyl subgroup. This means that arbitrary choices of  annihilation operators by means of the canonical Bogolyubov transformations [7] lead to unitary non-equivalent vacua. However, the embedding of the curved space $Q$ in a (centrally extended) group $\tilde{G}$ strongly restricts the possible Bogolyubov transformations [34]. It is also possible to choose particular states which behave as vacua with respect to a given subgroup $\tilde{G}_{K}\subset \tilde{G}$ i.e. the states which are invariant under $\tilde{G}_{K}$ only. The study of conformally invariant quantum field theories where Weyl (Poincare + dilatation) invariant pseudo-vacua (zero-mode coherent states) have been chosen is relevant for the Fulling-Unruh effect and vacuum radiation in relativistic accelerated frames [35]. Such a choice of particular pseudo-vacua corresponds to a symmetry breaking. References [33] and [34] also deal with the problem of second quantization from the perspective of the group approach. Indeed, as the choice of particular pseudo-vacua corresponds to a breakdown of the symmetry, it may result in a constrained version of our previous theory. At the second quantized level, we may, according to [34], introduce a corresponding constrained quantum field as well. The selection of a given Hilbert subspace $H^{\epsilon}(\tilde{G})\subset H(\tilde{G})$ made of wave functions obeying a higher-order constraint $K\psi=\epsilon\psi$ where $K$ is a Casimir operator of $\tilde{G}_{K}\subset\tilde{G}$ manifests itself at the level of second quantization in the form of a new quantum field theory. The vacuum for the new observables of this new (broken-symmetry) theory does not have to be identical to the vacuum of the original theory. The action of the other operators may produce vacuum radiation in this new theory. Such constructions have already shown how the Bogolioubov transformations can be restricted by demanding physical consistency and covariance to the arbitrary embeddings of spacetime in various groups. 

In this article I show how the choices of coefficient structures affect the sensitivity of the cohomology to more subtle cocycles. I showed in a previous article [9] how apparently trivial cohomology classes may actually be non-trivial when analyzed from the perspective of a different coefficient structure. An example of where this happens is the pseudo-cohomology, briefly presented in this article. 

I also study the topology of the groups involved by means of cohomology theories. The (co)homology with coefficients for the topology of manifolds with various dimensions has been studied for a long time. Relevant articles about the results and limitations of the associated methods are [26], [27], [28]. The analysis of groups by means of (co)homology is a somewhat newer idea, still presenting some interpretational gray areas. We know however that the second group cohomology of the group $G$ can be shown to be isomorphic with the set of associated extensions [36]. 
As the group extensions may characterize the vacua and the coefficient structures in cohomology control the extension by means of the universal coefficient theorem, it is possible to employ different coefficients in order to characterize different vacua of the same theory or different quantum field theories altogether, maybe obtained one from the other by breaking certain symmetries. The pairing between homology and cohomology induced by the universal coefficient theorem leads to topology-covariant constructions that may be transfered from one topological space to another. This may lead to an algebraic method for describing dualities in quantum field theories and string theory. 
The covariant formulation with respect to some transformations and the related ideas leading to equivalence principles (Galilei, Lorentz, Poincare) are important in this context. 
In particular, it is possible to relate the existence of a simple manifest covariant formulation and, in a more extended way, of an "equivalence principle" [4], to some topological properties of the transformation groups employed in the theory. 
The Unruh effect and the curved spacetime  in the group approach to quantization have been described in [33], [34]. 
The novelty of this article is the extension of these ideas for the situation when a topological covariant description would make the computations clearer. 

\section{2. Covariance principles in physics}
The main developments of the past century (special relativity, general relativity and quantum mechanics) have brought to our attention the fact that abstract mathematical conventions should not stand at the fundaments of a description of reality. In general, the role of conventions is to facilitate the comprehension of physical reality and not to assign physical reality to conventional constructions [3]. 
This statement can be translated in modern terminology by using (co)homological algebraic notations. In order to do this let me follow reference [4] and define
\begin{equation}
\mathcal{P}=Tr_{4}\circ L
\end{equation}
to be the Poincare group where $Tr_{4}$ is the four dimensional translation group and $L$ the Lorentz group and
\begin{equation}
G=Tr_{4}\circ L_{G}
\end{equation}
to be the Galilei group where again $Tr_{4}$ is the four dimensional translation group and $L_{G}$ is the group of galilean boosts and rotations. 
In contrast to the Poincare group, due to the absoluteness of time, the Galilei group admits several semi-direct structures. One can use for example the decomposition 
\begin{equation}
G=(((Tr_{3}\otimes B_{3})\circ T))\circ \mathcal{R}=H\circ \mathcal{R}
\end{equation}
where $Tr_{3}$ is the 3 dimensional translation group, $B_{3}$ is the 3 dimensional boost group, $T$ represents time translations and $\mathcal{R}$ represents rotations.
This allows one to define the mechanical evolution space as the homogeneous space parametrized by $(t,x,\dot{x})$.
This evolution space is however not a homogeneous space for the Poincare group, because of the different cohomological properties of the Galilei and Poincare groups: while $H_{0}^{2}(G,U(1))=R$ for the Galilei group, for the Poincare groups $H_{0}^{2}(\mathcal{P},U(1))=0$. This difference in the cohomological structures of the Galilei and Poincare groups has as consequence the absence of any simple covariant formulation of Newtonian mechanics, as opposed to the Poincare case [4]. In this way, the existence of a special topological structure of the symmetry group of a theory is related to the existence of a simple enough covariant formulation. This is not to say that a covariant formulation for the Newtonian mechanics is impossible. In fact, it is possible, after certain choices regarding the probing of topological properties are made. 
With this example I show that the topological features of the kinematical symmetry group are of utmost importance and hence the cohomology must be capable of distinguishing them even at an incipient level. This is translated in the requirement that the cohomology detects not only standard non-trivial cocycles but also pseudo-cocycles i.e. cocycles that would become non-trivial only when certain operations on the groups are being performed.
It is important to notice how this argument can be extended when one deals not only with covariance with respect to a symmetry group but with covariance to a change in the measurement technique for the topology of a group. 
Before entering this discussion it is important to put the observations above on firmer ground. Following reference [30] and [37] the Galilei group can in fact be obtained from the Poincare group by means of a group contraction when $c\rightarrow \infty$. Such a contraction gives rise to the notions of pseudo-cohomology and pseudo-extensions in the following way. Consider the trivial cohomology classes denoted by $[[\xi]]$. Inside these classes it is however sometimes possible to distinguish cohomology subclasses $[\xi]\in[[\xi]]$. These can be selected from the coboundaries according to some additional structure associated with the original group. Let me call now that original group $G$. The physical origin of pseudo-cohomology is as follows. We start from a given group $G$ for which we know a central extension $\tilde{G}$ associated with a two-cocycle $\xi_{cob}$ generated by a function $\lambda$ on $G$. Now, consider there exists a well defined contraction limit of the group $\tilde{G}$ producing $\tilde{G}_{c}$ in the sense of Inonu and Wigner. Therefore the two-cocycle $\xi_{cob}$ is well defined under the limit. It could however happen that the generating function $\lambda$ is ill-defined (divergent) in this limit. In that case the contracted two-cocycle is no longer a coboundary since there is no $\lambda_{c}$ to generate it. This procedure therefore was among the first to generate non-trivial group cohomology. It also shows that the triviality of cohomology depends on the limit behavior of the generators of the coboundary. The Lie algebra structure constant associated with a pseudo-extension, i.e. a central extension characterized by a pseudo-cocycle, therefore differs from that of the trivial product. This fact requires the non-triviality of the gradient of $\lambda$ at the identity of the group $G$. Pseudo-cocycles are generated by functions which are, locally, linear functions. For finite-dimensional semisimple groups for which the Whitehead lemma applies, pseudo-cohomology is important. In the case of infinite dimensional semisimple Lie groups for which the Whitehead lemma does not apply, the group law for $\tilde{G}$ will contain two-cocycles as well as pseudo-cocycles. 
The first physical example, connected to the discussion above appears in the contraction Poincare $\rightarrow$ Galileo where a special kind of trivial two-cocycles in the Poincare group become true two-cocycles for the Galilei group in the $c\rightarrow \infty$ limit. While the two-cocycle is well behaved in the limit, its generating function is not, initiating cohomology generation. 
Another simple physical example appears for the free non-relativistic particle with spin, where the Galilei group must be extended by a true two-cocycle to describe the canonical commutation relations between $q$ and $p$ as well as by a pseudo-cocycle associated with the Cartan subgroup of $SU(2)$ to account for the spin degree of freedom. 
Such global properties identifiable by means of pseudo-cohomology and pseudo-extensions are also identifiable by means of obstructions to the naive change of coefficient groups in cohomology.


\section{3. Independence of topology and the Universal Coefficient Theorem}
As argued in the previous chapters, the laws of physics should not depend on arbitrary choices. Specifically the choice of a particular coordinate system or a particular coefficient group in cohomology should not be relevant for the formulation of the laws of physics. I showed in a previous article [9] that specific choices of coefficient groups in cohomology may affect the observable connectedness of space-time (or generally of an abstract space or group) as measured by topological techniques. An interesting example for the role of the coefficient group in cohomology for the detection of topological properties is given in [31].  Here I focus on a different aspect, namely what changes should be made in a theory in order for it to describe the physical reality independent on the way one choses the coefficient structure in cohomology? 
As has been shown in [4] and as I argued in the previous sections, the existence of a trivial second group cohomology associated to a symmetry group implies the existence of a straightforward covariant formulation of the associated theory. 
The sensitivity of cohomology in a given dimension however, is controlled by the choice of a coefficient structure in the cohomology. The effect of this choice is on its turn, encoded in the universal coefficient theorem by means of the extensions. 
The quantization prescription in general and the form taken by the unitarity constraints depend on the topology of the space where quantization is performed. The visibility of the respective topological structures also depends on the coefficients in cohomology. Therefore, formulating the quantization with respect to a cohomology that is not sensitive to certain topological features (e.g. black hole horizons) may restore unitarity and a straightforward quantization. However, such a construction will have to be corrected by topology dependent terms. These will become manifest when universal coefficient theorems are being employed. These together with their extensions and torsions will play the topology analogue of differential forms in geometry. 


It has been brought as an argument for the information paradox that a relatively ordered initial situation (dust or a star) leading to a black hole has as an inescapable final state the thermal radiation. Unless some "emission of negative entropy" [1] by the black hole occurs, information should be lost. However, I showed in ref. [9] that the definition of entropy in a situation where different coefficient groups are required, must change. In fact, the entropy will have to include topological information as well. It will not be defined uniquely. Instead it will have different forms when regarded via different coefficient groups. This allows the changes in entropy required to restore unitarity in a global (topological) way. 

As stated in the previous chapters, geometric quantization is tightly bound to the existence of a classical limit. In the absence of such a limit the associated methods are often insufficient. In order to circumvent this problem a new quantization method has been developed [32] based on a group theoretical approach. The references [29] and [30] are also relevant for a better understanding of this. The main ingredient of such a method is a Lie group structure on the manifold replacing the quantum manifold of geometric quantization. This Lie group, which I call $\tilde{G}$ also appears to be a principal bundle with structure group $U(1)$. However, in this more general approach $\tilde{G}/U(1)$ is not forced to have a symplectic structure. Therefore non-symplectic symmetry parameters are naturally allowed giving rise to the corresponding operators (Hamiltonian, angular momentum, null charges, etc.). Another advantage is that on any Lie group there are always two sets of mutually commuting vector fields. The group approach to quantization is not meant to quantize a classical system (a phase space) but instead, the quantizing group is the primary quantity. 

The group approach to quantization is a more general approach based on the quantizing group $\tilde{G}$ with a principal bundle structure $\tilde{G}(M,T)$ having $T$ as a structure group and $M$ its base. The group $T$ generalizes the phase invariance of quantum mechanics. It will encode constraints that may introduce topological properties that have to be taken into account. Of course the simplest yet general case remains $T=U(1)$. The group law for $\tilde{G}=\{\tilde{g}=(g,\zeta)/g\in G, \zeta\in U(1)\}$ is
\begin{equation}
\tilde{g}'*\tilde{g}=(g'*g, \zeta'\zeta e^{i\xi(g',g)})
\end{equation}
the group operation in $G$ being $g"=g'*g$ and $\zeta(g',g)$ is a two co-cycle of $G$ with the property that
\begin{equation}
\xi(g_{2},g_{1})+\xi(g_{2}*g_{1},g_{3})=\xi(g_{2},g_{1}*g_{3})+\xi(g_{1},g_{3}),\; g_{i}\in G
\end{equation}
We say the central extensions are trivial if the two co-cycles are coboundaries which can be written in the form of 
\begin{equation}
\xi(g',g)=\delta(g'*g)-\delta(g')-\delta(g)
\end{equation}
where $\delta(g)$ is the generating function of the co-boundary. The group $\tilde{G}$ acting on itself on the right and on the left provide two sets of mutually commuting invariant vector fields 
\begin{equation}
\tilde{X}^{L}_{\tilde{g}^{i}}=\frac{\partial \tilde{g}"^{j}}{\partial \tilde{g}^{i}}\biggr\rvert_{\tilde{g}=e}\frac{\partial}{\partial \tilde{g}^{j}}
\end{equation}
\begin{equation}
\tilde{X}^{R}_{\tilde{g}^{i}}=\frac{\partial \tilde{g}"^{j}}{\partial \tilde{g}^{'i}}\biggr\rvert_{\tilde{g'}=e}\frac{\partial}{\partial \tilde{g}^{j}}
\end{equation}
\begin{equation}
[\tilde{X}^{L}_{\tilde{g}^{i}},\tilde{X}^{R}_{\tilde{g}^{j}}]=0
\end{equation}
where $\{\tilde{g}^{j}\}$ is a parametrization of $\tilde{G}$. Next, the left invariant quantization $1$-form $\Theta$ associated with the central generator $\tilde{X}^{L}_{\zeta}=\tilde{X}^{R}_{\zeta},\; \zeta\in T$ namely the $T$-component of $\tilde{\theta}^{L(\zeta)}$ of the canonical left invariant $1$-form $\tilde{\theta}^{L}$ on $\tilde{G}$. The differential $d\Theta$ is a presymplectic form and its characteristic module $Ker(\Theta)\cap Ker(d\Theta)$ is generated by a left subalgebra $\mathcal{G}_{\Theta}$. The quotient group $(\tilde{G},\Theta)/\mathcal{G}_{\Theta}$ is a quantum manifold. The trajectories generated by the vector fields in $\mathcal{G}_{\Theta}$ are the generalized equations of motion of the theory. The Noether invariants under those equations are $F_{\tilde{g}^{j}}=i_{\tilde{X}_{\tilde{g}^{j}}^{R}}\Theta$. One may consider the set of complex-valued $T$-functions on $\tilde{G}$ in the sense of principal bundle theory: 
\begin{equation}
\psi(\zeta*\tilde{g})=D_{T}(\zeta)\psi(\tilde{g}),\; \zeta\in T
\end{equation}
where $D_{T}$ is the natural representation of $T$ on the complex numbers. The representation of $\tilde{G}$ on the set of complex valued $T$-functions generated by $\mathcal{G}^{R}=\{\tilde{X}^{R}\}$ is called Bohr quantization. However, this quantization is as it stands, reducible. To obtain an irreducible representation one may impose a full polarization $\mathcal{P}$
\begin{equation}
\tilde{X}^{L}\psi_{p}=0,\; \forall \tilde{X}^{L}\in \mathcal{P}
\end{equation}
which is a maximal, horizontal left subalgebra of $\tilde{\mathcal{G}}^{L}$ which contains $\mathcal{G}_{\Theta}$. The existence of a full polarization for the whole subalgebra $\mathcal{G}_{\Theta}$ is not guaranteed. In this case a higher order polarization will be required which is a polarization for the enveloping algebra $U\mathcal{G}^{L}$ which contains $\mathcal{G}_{\Theta}$. This higher order polarization will be made of the extended vector fields corresponding to the momentum variables, spacetime symmetries and internal symmetries together with a deformation of the vector field $\tilde{X}_{t}^{L}$ associated with the temporal evolution which, usually, can be chosen to be the Casimir operator of $G$
\begin{equation}
\mathcal{P}^{HO}=<\tilde{X}^{HO}_{t},\tilde{X}^{L}_{h^{i}}>
\end{equation}
The group $\tilde{G}$ can be irreducibly represented on the space $\mathcal{H}(\tilde{G})=\{\ket{\psi}\}$ of polarized wavefunctions and on its dual $\mathcal{H}^{*}(\tilde{G})=\{\bra{\psi}\}$. The coordinates of the ket and the bra in the representation defined through the polarization $\mathcal{P}$ are
\begin{equation}
\begin{array}{c}
\psi_{p}(\tilde{g})=<\tilde{g}_{\mathcal{P}}|\psi>\\
\psi^{'*}_{p}(\tilde{g})=<\psi'|\tilde{g}_{\mathcal{P}}>\\
\end{array}
\end{equation}
This leads to the inner product of the form 
\begin{equation}
<\psi'|\psi>=\int_{\tilde{G}}\mu(\tilde{g})\psi^{'*}_{\mathcal{P}}(\tilde{g})\psi_{\mathcal{P}}(\tilde{g})
\end{equation}
with 
\begin{equation}
\mu(\tilde{g})=\theta_{\tilde{g}^{i}}^{L}\wedge \theta_{\tilde{g}^{j}}^{L}\wedge ... 
\end{equation}
The closure relation in $\tilde{G}$ is 
\begin{equation}
1=\int_{\tilde{G}}\ket{\tilde{g}_{\mathcal{P}}}\mu(\tilde{g})\bra{\tilde{g}_{\mathcal{P}}}
\end{equation}
The group $\tilde{G}$ has a unitary representation $\rho$ such that 
\begin{equation}
\bra{\tilde{g}_{\mathcal{P}}}\rho(\tilde{g}')\ket{\psi}=\psi_{\mathcal{P}}(\tilde{g}^{'-1}*\tilde{g})
\end{equation}
Enlarging the the structure group $T$ allows us to introduce constraints in the theory. These constraints should always include $U(1)$. The constraints are being introduced by means of the $T$-function conditions 
\begin{equation}
\rho(\tilde{t})\ket{\psi}=D_{T}^{(\epsilon)}(\tilde{t})\ket{\psi},\; \tilde{t}\in T
\end{equation}
For example quantum mechanics on a non-simply connected manifold $Q$ may be recovered from quantum mechanics on its universal covering $\bar{Q}$ by choosing $T=\pi_{1}(Q)\otimes U(1)$ as the structure group. This is well known and leads to the so called topological quantum effects known as the $\theta$-structure. 
If the structure group $T$ is non-central, not all operators $\tilde{X}_{\tilde{g}}^{R}$ preserve the constraints imposed within $T$. For this to happen for a subgroup $\tilde{G}_{T}\subset \tilde{G}$ we need that 
\begin{equation}
[\tilde{G}_{T},T]\subset Ker(D_{T}^{\epsilon})
\end{equation}
The classes of inequivalent two-cocycles for the topologically trivial quantization define the second cohomology group $H^{2}(G;U(1))$. However, as showed above (and also following from [32]), in order to introduce constraints which may imply non-trivial topological structure the group $U(1)$ may be replaced with a more general structure. This would lead to a construction of the type $H^{2}(G;T)$ which expands on the standard cohomology and replaces the usual coefficient structure with another one, capable to detect additional topological structure. Modified coefficient groups therefore allow the precise choice of the level of refinement demanded from a particular cohomology theory.

However, there exist physical properties which are basically independent of a particular topology. Dualities relating open and closed strings are relevant examples. In order to construct such topology-independent theories, it appears to be necessary to employ the universal coefficient theorems. These theorems state that a specific framework, constructed by the choice of a coefficient group in (co)homology is (up to (extension) torsion in (co)homology) equivalent with the choice of an integer coefficient group. 
One result of this theorem is that distinct classes in (co)homology under one coefficient group may appear as identified under another coefficient group. 

There are several ways in which we can generalize the usual pairings relating vectors and 1-forms to pairings relating homology and cohomology. Such more general pairings will involve the universal coefficient theorems and the required covariance will be translated into rules relating the possible extensions arising in the associated exact sequence.

One such possible pairing is defined as
\begin{equation}
< , >:H^{q}(G;M)\times H_{q}(G)\rightarrow M
\end{equation}
which relates homology with cohomology. 
This pairing is bilinear and its adjoint is a homomorphism 
\begin{equation}
H^{q}(G;M)\rightarrow Hom(H_{q}(G);M)
\end{equation}
Universal coefficient theorems, among other things, provide a measure of how this adjoint fails to be an isomorphism in terms of $Ext$ and $Tor$ [10]. Here $q$ represents the dimension of the space for which the (co)homology is calculated.

In the context of the group approach to quantization the main topological tool to be used is group (co)homology. This cohomology measures the extent to which the invariants of the groups in an exact sequence do not respect the original exact sequence. Consider for example an abelian group $M$ together with a group action of the group $G$ on $M$. The elements of $G$ are acting as an automorphism of $M$. One may consider the submodule of $G$-invariant elements of $M$ 
\begin{equation}
M^{G}=\{x\in M | \forall g\in G : gx=x\}
\end{equation}
Consider $N$ as a $G$-submodule of $M$. The invariants in $M/N$ are in general not the quotient of the invariants in $M$ by the invariants in $N$. The invariance modulo $N$ is a broader concept, measured precisely by the first group cohomology $H^{1}(G,N)$. In order to describe the homology groups for trivial group action of $G$ on $M$ in terms of the homology groups for trivial group action of $G$ on a reference group, say $\mathbb{Z}$ one may use the universal coefficient theorem for group homology
\begin{equation}
0\rightarrow H_{p}(G;\mathbb{Z})\otimes M \rightarrow H_{p}(G;M)\rightarrow Tor(H_{p-1}(G;\mathbb{Z}),M)\rightarrow 0
\end{equation}
The sequence splits, although not naturally giving 
\begin{equation}
H_{p}(G;M)\cong (H_{p}(G;\mathbb{Z})\otimes M)\oplus Tor(H_{p-1}(G;\mathbb{Z}),M)
\end{equation}
However, to connect homology with cohomology in a way useful to the study of how the scalar products change at a change of coefficients a more useful result is that of the dual universal coefficient theorem, linking the homology groups for trivial group action of $G$ on $\mathbb{Z}$ and the cohomology group for trivial action of $G$ on $M$ 
\begin{equation}
0\rightarrow Ext(H_{p-1}(G;\mathbb{Z}),M)\rightarrow H^{p}(G;M)\rightarrow Hom(H_{p}(G;\mathbb{Z}),M)\rightarrow 0
\end{equation}
Here as well, the sequence splits although not naturally 
\begin{equation}
H^{p}(G;M)\cong Hom(H_{p}(G;\mathbb{Z})\oplus Ext(H_{p-1}(G;\mathbb{Z}),M)
\end{equation}
In the case of second order cohomology, with $G$ again a group and $A$ an abelian group we have 
\begin{equation}
0\rightarrow Ext^{1}(G^{ab},A)\rightarrow H^{2}(G;A)\rightarrow Hom(H_{2}(G;\mathbb{Z}),A)\rightarrow 0
\end{equation}
where $Hom(H_{2}(G;\mathbb{Z}),A)$ called the second cohomology group up to isoclinism is the group of group homomorphisms from $H_{2}(G;\mathbb{Z})$ to $A$. The group $H_{2}(G;\mathbb{Z})$ is the second homology group for the trivial group action. The extension group $Ext^{1}(G^{ab},A)$ describes abelian group extensions with normal subgroup $A$ and quotient group $G^{ab}$. The map $Ext^{1}(G^{ab},A)\rightarrow H^{2}(G;A)$ leads us from the abelian group extension with normal subgroup $A$ and quotient group $G^{ab}$ to the extension of $G$ by $A$. 
A short exact sequence of groups given as
\begin{equation}
0\rightarrow A \rightarrow E \rightarrow G\rightarrow 1
\end{equation}
with $E$ the central extension, meaning that the image of $A$ in $E$ is a central subgroup of $E$, defines a natural homomorphism $\beta:M(G)\rightarrow A$ with $M(G)=H_{2}(G;\mathbb{Z})$. 
Suppose then that we fix the abelian group $A$ and the group $G$ and let the central extension group $E$ vary. The structures that $E$ can take up to congruence correspond to $H^{2}(G;A)$. For each extension we obtain an element of $Hom(M(G),A)$. Congruent group extensions define the same homomorphism. Therefore we obtain our homomorphism $H^{2}(G;A)\rightarrow Hom(M(G),A)$ which is surjective. Therefore, the universal coefficient theorem appears to be a tool for identifying the changes in the extensions. Moreover, it can be interpreted as a pairing, therefore demanding particular multipliers originating in the various $Ext$ groups appearing on the left, which practically encode non-trivial two-cocycles. These will be the multiplicative elements arising in the scalar product formulas and leading to non-thermal corrections to the Hawking radiation.

\section{4. Black Holes and the unitarity problem}
The previous sections showed that when using equivalence principles and covariant formulations of theories, one usually relies on specific topological properties of the symmetry groups. Especially the second group-cohomology, when trivial, allows for a simple covariant formulation as the one used in the bra-ket formalism or in the tensorial construction of general relativity. 
However, not in all situations is the second group-cohomology trivial. The sensitivity of the second cohomology group to various topological features (non-trivial cocycles, pseudo-cocycles, etc.) depends on the coefficient structure chosen in order to describe the cohomology itself. 
When the second cohomology of the required group is non-trivial one can still formulate a covariant theory provided one uses a proper coefficient structure, giving the desired sensitivity to the cohomology groups. The universal coefficient theorem applied to group cohomology proves to be useful in analyzing what happens when the coefficient structure is changed. 

In this section, I present some physical arguments for the necessity of a coefficient independent construction and, implicitly, of theories that do not depend on how precisely cohomology can probe certain topological features. If in the case of general relativity and quantum mechanics the covariance had to be implemented with respect to a symmetry group, in order to implement the topological covariance one has to consider the coefficient structures in (co)homology and the associated extensions. 
Probably the most important object for which the current discussion is relevant is a black hole. The problem of information conservation was discussed in the context of quantized fields over a given background in [1]. I partially follow the discussion presented therein, pinpointing the aspects where an extension of that treatment is necessary due to some ignored topological aspects. 
Considering, in agreement with [1] a massless Hermitian scalar field and an uncharged non-rotating black hole, after quantization one obtains a scalar field operator $\phi$ which satisfies the wave equation 
\begin{equation}
\Box \phi = 0
\end{equation}
Given the background metric associated to the Schwarzschild spacetime [5] where the considered black hole is present one can rewrite this as
\begin{equation}
(-g)^{\frac{1}{2}}\partial_{\mu}[(-g)^{\frac{1}{2}}g^{\mu\nu}\partial_{\nu}\phi]=0
\end{equation}
One can also define a conserved scalar product of the form 
\begin{equation}
(\phi_{1},\phi_{2})=i\int d^{n-1}x|g|^{1/2}g^{0\nu}\phi_{1}^{*}(x,t)\overleftrightarrow{\partial_{\nu}}\phi_{2}(x,t)
\end{equation}
the integral being over a constant $t$ hypersurface. When $\phi_{1}$ and $\phi_{2}$ are solutions of the field equation above and vanish at spatial infinity, then $(\phi_{1},\phi_{2})$ is conserved. 
The existence of a flow of particles originating at a small affine distance from the event horizon has been derived in [2].  One particularity of this derivation is that the average number of outgoing particles in each mode is distributed in accordance with a thermal spectrum. Moreover, the full probability distribution, not just the average, of the emitted particles is that of thermal radiation. 
This observation creates a conflict with standard quantum mechanics when one considers the process of an in-falling object together with the radiation emitted on the external part of the horizon. The main issue is that this process does not preserve unitarity. If the in-falling system is in a pure quantum state, the out-coming radiation is in a naturally mixed state. The full information related to the in-falling object is forever hidden behind the horizon. This result, however, appears only when one does not consider the process as described in a topologically covariant way. 

Using some of the observations in [9] I show here that there exists a special choice of coefficients in the quantization group cohomology for which the quantization prescription (particularly the group approach to quantization) allows a unitary connection between the outgoing radiation and the in-falling system. This suggests that the quantum information is in fact conserved, albeit not in the obvious way, but instead in a way visible only by means of cohomology with carefully chosen coefficients. 

In order to show this I continue the derivation of the spectrum of the Hawking radiation underlining the modifications in the way of thinking that must be considered in order to obtain the correct result. This method is in agreement with the AdS/CFT solution but its construction allows for a higher degree of generality. 
Let me now take the quantum fields used in the field equation above and decompose them as 
\begin{equation}
\phi = \int d\omega (a_{\omega}f_{\omega}+a_{\omega}^{+}f_{\omega}^{*})
\end{equation}
where $f_{\omega}$ and $f_{\omega}^{*}$ form a complete set of solutions of the field equation and are normalized according to 
\begin{equation}
(f_{\omega_{1}},f_{\omega_{2}})=\delta(\omega_{1}-\omega_{2})
\end{equation}

The $a_{\omega}$ operators are time independent. 
The standard method of quantization (second quantization) would be 
\begin{equation}
\begin{array}{c}
[ a_{\omega_{1}}, a_{\omega_{2}}^{+} ] =\delta(\omega_{1}-\omega_{2}) \\
\\
0 =  [a_{\omega_{1}}^{+}, a_{\omega_{2}}^{+}] = [a_{\omega_{1}}, a_{\omega_{2}}] \\
\end{array}
\end{equation}
Let me chose the $f_{\omega}$ such that at early times and large distances they form a complete set for the incoming positive frequency solutions of energy $\omega$. It is possible to compute the spectrum of the created particles by making an expansion of the field in terms of the late time positive frequency solutions. Let $p_{\omega}$ be the solutions of the field equation that have zero Cauchy data on the event horizon and are asymptotically out-coming with positive frequency. 
Again, consider that in this domain $p_{\omega}$ and $p_{\omega}^{*}$ form a complete set of solutions. 
The normalization condition is 
\begin{equation}
(p_{\omega_{1}},p_{\omega_{2}})=\delta(\omega_{1}-\omega_{2})
\end{equation}
There must also be an in-coming component of the solution at the event horizon at late times. Let me call this set of solutions $q_{\omega}$. The superposition of these components at late times is localized on the horizon and has zero Cauchy data on the distant region. The components $q_{\omega}$ and $q_{\omega}^{*}$ form a complete set on the horizon and are normalized as
\begin{equation}
(q_{\omega_{1}},q_{\omega_{2}})=\delta(\omega_{1}-\omega_{2})
\end{equation}
The two components, being defined in disjoint regions are assumed to have null scalar product
\begin{equation}
(q_{\omega_{1}},p_{\omega_{2}})=0
\end{equation}
The expansion of the fields in terms of the above components is then 
\begin{equation}
\phi=\int d\omega \{ b_{\omega}p_{\omega}+c_{\omega}q_{\omega}+b_{\omega}^{+}p_{\omega}^{*}+c_{\omega}^{+}q_{\omega}^{*}\}
\end{equation}
where $b_{\omega}$ and $c_{\omega}$ are the associated annihilation operators. The commutation relations are now
\begin{equation}
\begin{array}{c}
[b_{\omega_{1}},b_{\omega_{2}}^{+}]=\delta(\omega_{1}-\omega_{2})\\

\\

[c_{\omega_{1}},c_{\omega_{2}}^{+}]=\delta(\omega_{1}-\omega_{2})\\
\end{array}
\end{equation}
all other commutators are vanishing. 
The spectrum of the outgoing particles is determined by the coefficients of the Bogolubov transformation relating $b_{\omega}$ to $a_{\omega'}$ and $a_{\omega'}^{+}$.
One may define the operators $c_{\omega}$ and $c_{\omega}^{+}$ as the annihilation and creation operators for particles falling into the black hole. However, this definition is ambiguous due to the fact that the positive frequency components for the in-falling matter are not well defined. The physical meaning of these operators should therefore be taken as symbolic. 
Using the complete set given by $f_{\omega}$ and $f_{\omega}^{*}$ one can write 
\begin{equation}
p_{\omega}=\int d\omega' (\alpha_{\omega\omega'} f_{\omega'}+\beta_{\omega\omega'} f_{\omega'}^{*})
\end{equation}
where $\alpha$ and $\beta$ are complex numbers, independent of the coordinates. 
We can therefore calculate 
\begin{equation}
b_{\omega}=(p_{\omega},\phi)
\end{equation}
and expressing $\phi$ and $p_{\omega}$ in terms of  $f_{\omega'}$ and $f_{\omega'}^{*}$ one can obtain 
\begin{equation}
b_{\omega}=\int d\omega(\alpha_{\omega\omega'}^{*}a_{\omega'}-\beta_{\omega\omega'}^{*}a_{\omega'}^{+})
\end{equation}
and the invariant becomes 
\begin{equation}
(p_{\omega_{1}},p_{\omega_{2}})=\int d\omega'(\alpha^{*}_{\omega_{1}\omega'}\alpha_{\omega_{2}\omega'}-\beta^{*}_{\omega_{1}\omega'}\beta_{\omega_{2}\omega'})
\end{equation}
It is worthwhile noticing that the coefficients can be expressed as 
\begin{equation}
\begin{array}{c}
\beta_{\omega\omega'}=-(f_{\omega'}^{*},p_{\omega})\\

\\

\alpha_{\omega\omega'}=(f_{\omega'},p_{\omega})\\
\end{array}
\end{equation}


The discussion up to this point is unsurprising. The calculation of the coefficients above can be used in order to derive the average number of created particles observed at later times. However the exact form in which the previous calculations are being performed does not take the fact into account that the topology as encoded by cohomology groups changes when a black hole forms. Moreover, it is not clear that the coefficient groups in the associated cohomology are good enough to take into account the new topology. While the curvature of spacetime is correctly taken into account in the previous discussion, there are certain modifications required for the pairings to be isomorphically translated from the language of flat or curved spacetime to the language of spacetime with a horizon. 

It is important to notice that there are several possible choices of topologies over a space. One possible choice would be to consider any two points joined together in a subset for a specific topology if they can be connected by light in both directions. The space filled with low density dust before the formation of a black hole has every point connected in such a topology. Once a horizon forms the topology defined in the above way changes. Moreover, after the horizon is formed, any topology that, prior to the formation of the horizon, connected two points on different sides of what is now the horizon, must change in order to consider the new situation. Such a choice of topology would naturally incorporate causality. Various choices of topology have been discussed in [11-17]. For a discussion of homology with non-trivial coefficients and suggestive examples of the applications of the universal coefficient theorem ref. [18-21] could be relevant for the reader. For discussions on topology changes references [22-25] are recommended.


\par Because of this change of topology, each of the constructions defined above has to be carefully analyzed. For this I will employ a group approach to quantization and adapt this language to the second quantization prescription. In doing this I mainly follow reference [34]. However, already at the level of the Poincare group, we face a problem when dealing with relativistic quantum mechanics. There appears to be no position operator $\hat{x}$ satisfying the commutation relation $[\hat{x},\hat{p}]=i\hbar \hat{1}$ with the ordinary momentum operator $\hat{p}$. The group approach to quantization is formulated in terms of the extended group $\tilde{G}$ which is a principal bundle with fiber $U(1)$. In the case of geometric quantization one requires a symplectic form $\omega$ and the existence of a polarization i.e. a maximal (half the dimension of the manifold) isotropic distribution of vector fields with respect to the symplectic form. When such a polarization does not exist, geometric quantization cannot be performed. In the group approach, the symplectic form is replaced by $d\Theta$ where $\Theta$ is the left $1$-form dual to the vertical generator (the generator tangent to the fiber). The non-polarizability of the geometric quantization is translated here into the absence of a first order full polarization i.e. a maximal left subalgebra containing $Ker(\Theta)$ and excluding the $U(1)$-generator. However, here, this anomaly can be avoided by adding additional operators in the left enveloping algebra to a non-full first order polarization and so defining a higher order polarization. Moreover, higher-order polarizations have been shown to be useful for representing a physical system in an equivalent but different realization than that given by the first order full polarizations.

In the group approach to quantization, as the whole spacetime is embedded into a larger group structure containing the phase space of the problem, the group cohomology becomes a key element in identifying possible obstructions to quantization. Particularly, a situation not considered up to now is that in the process of black hole formation, the cohomology must be sensitive enough to detect the formation of a horizon. Therefore, whatever coefficient group was employed before the formation of the black hole, thereafter we need a coefficient group that gives sufficient sensitivity to the cohomology and allows a proper quantization by means of the group approach. Therefore, a set of requirements arise. First, a universal coefficient theorem for group cohomology must exist, which provides us with the properties that remain unchanged when a topology change occurs. It also gives us via $Ext$ or $Tor$ groups the properties that must change during such a process (mainly corrections to quantization prescriptions). Such a universal formula exists in general (although particular exceptions may be found) only for trivial group action of the considered group on the coefficient group. Moreover, the coefficients are being associated in equivalence classes as dictated by the morphisms between the resulting extensions. Isomorphic group extensions will correspond to equivalent coefficient groups. Finally, demanding that the pairing resulting in the scalar product of fields and including the homology-cohomology pairing is covariant with respect to the change in topology implies certain corrections to the quantization prescription.
 Let me therefore, before discussing the effects of changing coefficients in (co)homology perform the quantization of a free relativistic particle by means of the group approach following mainly [34]. For the sake of simplicity, I will work in the $(1+1)$-dimensional Minkowski spacetime parametrized by $\{a^{\mu}\}=\{a^{0},a^{1}=a\}$ where the transformations are given by 
\begin{equation}
a'^{\mu}=\Lambda^{\mu}_{\;\nu}(p^{0},p)a^{\nu}+x^{\mu}
\end{equation}
with $\{x^{\mu}\}$ the translations and $\{\Lambda\}$ the boosts. As a manifold, the group can be seen as the direct product of Minkowski spacetime and the mass hyperboloid. The composition law of the group is given by two consecutive actions of Poincare transformations
\begin{equation}
\begin{array}{c}
x^{0"}=x^{0'}+\frac{p^{0'}}{mc}x^{0}+\frac{p'}{mc}x\\
\\
x"=x'+\frac{p^{0'}}{mc}x+\frac{p'}{mc}x^{0}\\
\\
p"=\frac{p^{0}}{mc}p'+\frac{p^{0'}}{mc}p\\
\\
\end{array}
\end{equation}
It is known that the Poincare group admits only trivial extensions by $U(1)$ i.e. extensions of the form 
\begin{equation}
\tilde{g}'*\tilde{g}=(g'*g,\zeta'\zeta e^{i\xi(g',g)})
\end{equation}
where the cocycle $\xi$ is a coboundary generated by a function on $G$ satisfying either $\gamma^{-1}(g'g)\cdot \gamma(g')\cdot\gamma(g)$ or in the case of an additive group (the exponents of the law) $\xi(g',g)=\delta(g'*g)-\delta(g')-\delta(g)$. We choose $\delta(g)=mcx^{0}$ so that the law to be added to the laws of the Poincare group should be
\begin{equation}
\zeta"=\zeta' \zeta e^{imc(x^{0"}-x^{0'}-x^{0})}
\end{equation}
The left and right invariant fields are to be calculated according to the law 
\begin{equation}
\begin{array}{c}
\tilde{X}^{L}_{\tilde{g}^{i}}=\frac{\partial \tilde{g}^{"j}}{\partial \tilde{g}}\biggr\rvert_{\tilde{g}=e} \frac{\partial}{\partial \tilde{g}^{j}}\\
\tilde{X}_{\tilde{g}'^{i}}^{R}=\frac{\partial\tilde{g}"^{j}}{\partial \tilde{g}'^{i}}\biggr\rvert_{\tilde{g}'=e} \frac{\partial}{\partial \tilde{g}^{j}}\\
\end{array}
\end{equation}
satisfying the commutation relations
\begin{equation}
[\tilde{X}_{\tilde{g}^{i}}^{L},\tilde{X}_{\tilde{g}^{j}}^{R}]=0
\end{equation}
Applied for the free relativistic particle these invariant fields become
\begin{equation}
\begin{array}{c}
\tilde{X}_{x_{0}}^{L}=\frac{p^{0}}{mc}\frac{\partial}{\partial x^{0}}+\frac{p}{mc}\frac{\partial}{\partial x}+\frac{p^{0}(p^{0}-mc)}{mc}i\zeta\frac{\partial}{\partial \zeta}\\
\\
\tilde{X}_{x}^{L}=\frac{p^{0}}{mc}\frac{\partial}{\partial x}+\frac{p}{mc}\frac{\partial}{\partial x^{0}}+\frac{p(p^{0}-mc)}{mc}i\zeta\frac{\partial}{\partial \zeta}\\
\\
\tilde{X}_{p}^{L}=\frac{p^{0}}{mc}\frac{\partial}{\partial p}+\frac{p^{0}}{mc}x i  \zeta \frac{\partial}{\partial \zeta}\\
\\
\tilde{X}_{\zeta}^{L}=\tilde{X}_{\zeta}^{R}=\zeta\frac{\partial}{\partial\zeta}\\
\\
\tilde{X}_{x^{0}}^{R}=\frac{\partial}{\partial x^{0}}\\
\\
\tilde{X}_{x}^{R}=\frac{\partial}{\partial x}+p i \zeta \frac{\partial}{\partial\zeta}\\
\\
\tilde{X}_{p}^{R}=\frac{p^{0}}{mc}\frac{\partial}{\partial p}+\frac{x^{0}}{mc}\frac{\partial}{\partial x}+\frac{x}{mc}\frac{\partial}{\partial x^{0}}+(\frac{p}{mc}x^{0}+\frac{p^{0}}{mc}x-x) i \zeta \frac{\partial}{\partial \zeta}\\
\\
\end{array}
\end{equation}
Next, the pseudo-extended Poincare algebra is defined by 
\begin{equation}
[\tilde{X}_{x_{0}}^{R},\tilde{X}_{x}^{R}]=0
\end{equation}
\begin{equation}
[\tilde{X}_{x_{0}}^{R},\tilde{X}_{p}^{R}]=\frac{1}{mc}\tilde{X}_{x}^{R}
\end{equation}
\begin{equation}
[\tilde{X}_{x}^{R},\tilde{X}_{p}^{R}]=\frac{1}{mc}\tilde{X}_{x^{0}}^{R}-i\tilde{X}_{\zeta}^{R}
\end{equation}
The quantization $1$-form and the Characteristic subalgebra are 
\begin{equation}
\begin{array}{c}
\Theta=-(p^{0}-mc)dx^{0}-xdp-i\zeta^{-1}d\zeta\\
\\
\mathcal{G}_{\Theta}=<\tilde{X}_{x^{0}}^{L}>\\
\end{array}
\end{equation}
We define the characteristic module as $Ker(\Theta)\cap Ker(d\Theta)$. It is in general generated by the left subalgebra $\mathcal{G}_{\Theta}$ which we call the characteristic subalgebra. The trajectories generated by the vector fields in $\mathcal{G}_{\Theta}$ constitute the generalized equations of motion of the theory. 
The pseudo-extended Poincare group admits a first order full polarization 
\begin{equation}
\mathcal{P}_{p}=<\tilde{X}^{L}_{x^{0}},\tilde{X}^{L}_{x}>
\end{equation}
which generates the momentum representation. The polarized $U(1)$-functions satisfying $\tilde{X}_{\zeta}^{R}\psi=\psi$ are 
\begin{equation}
\psi_{\mathcal{P}_{p}}=\zeta exp[-i(p^{0}-mc)x^{0}]\phi(p)
\end{equation}
and the right generators act on them as quantum operators. 
This generates a representation which is unitary considering the measure of integration restricted to the momentum space
\begin{equation}
\mu(\tilde{g})=-i\frac{mc}{p^{0}}dx^{0}\wedge dx\wedge dp \wedge \zeta^{-1}d\zeta\rightarrow i_{\tilde{X}^{L}_{\zeta}}i_{\tilde{X}^{L}_{x}}i_{\tilde{X}^{L}_{x^{0}}}\mu(\tilde{g})=\frac{mc}{p^{0}}dp
\end{equation}
where $i_{X}$ induces the contraction of the invariant vector field $X$ with $\mu(\tilde{g})$. 
On the other side if we want to arrive at the configuration representation we notice there is no first order full polarization leading to it. The vector fields $\tilde{X}^{L}_{p}$ and $\tilde{X}^{L}_{x^{0}}$ do not close a proper horizontal subalgebra. In order to avoid this, we can go to higher order polarizations that also include the $\tilde{X}^{L}_{p}$ generator
\begin{equation}
\mathcal{P}_{x}^{HO}=<\tilde{X}_{x_{0}}^{L\, HO}=\tilde{X}_{x^{0}}^{L}+i\cdot[\sqrt{m^{2}c^{2}-(\tilde{X}_{x}^{L})^{2}}-mc]\tilde{X}^{L}_{\zeta},\tilde{X}_{p}^{L}>
\end{equation}
After solving the polarization equations for this polarization we arrive at the following expression of the wavefunctions
\begin{equation}
\psi_{\mathcal{P}_{x}}=\zeta exp(-ixp)exp\{-imcx^{0}(\sqrt{m^{2}c^{2}-\frac{\partial^{2}}{\partial x^{2}}}-mc)\}\phi(x)
\end{equation}
The right invariant vector fields preserve this form of the polarized wavefunction and therefore we generally operate only on the functions $\phi(x)$. The resulting realization restricted to $(x,\frac{\partial}{\partial x})$ is not unitary with the trivially regularized restriction of the measure 
\begin{equation}
\mu(\tilde{g})\rightarrow dx
\end{equation}
although the representation is unitary on the complete wavefunctions. This failure is a consequence of the weak closure of the higher order polarization. While it closes on polarized wavefunctions giving a well defined irreducible carrier subspace, the polarization itself is not integrable in the classical sense. The transverse space to the polarization subalgebra which should be the $x$-space is not properly defined in the weak case. By performing the transformation 
\begin{equation}
\phi(x)=e^{-ix^{0}\hat{p}^{0}}\sqrt{\hat{p}^{0}}\varphi(x,x^{0})
\end{equation}
we regain unitarity, taking the boost operators in the symmetrized form 
\begin{equation}
\frac{1}{\sqrt{\hat{p}^{0}}}\hat{k}\sqrt{\hat{p}^{0}}=\frac{1}{2}(\frac{x}{mc}\sqrt{m^{2}c^{2}-\frac{\partial^{2}}{\partial x^{2}}}+\sqrt{m^{2}c^{2}-\frac{\partial^{2}}{\partial x^{2}}}\frac{x}{mc})
\end{equation}
which agrees with the standard prescription for the scalar product of relativistic fields 
\begin{equation}
\int dx \phi^{*}(x)\phi'(x)=\frac{1}{2}\int dx \varphi^{*}(x,x^{0})\overleftrightarrow{\partial_{0}}\varphi'(x,x^{0})
\end{equation}

In the case of general spacetime the scalar product receives a well known correction to the measure 
\begin{equation}
(\phi_{1},\phi_{2})=i\int d^{n-1}x|g|^{1/2}g^{0\nu}\varphi_{1}^{*}(x,t)\overleftrightarrow{\partial_{\nu}}\varphi_{2}(x,t)
\end{equation}
Such a scalar product depends on the topology of space at least at the level of the second cohomology. 
Up to this point we have seen how the group approach to quantization is performed when the topology is trivial. However, in order to develop a language capable of identifying the topology invariant properties and to transfer them from one topology to the other, a discussion about non-trivial topologies is required. As we have seen previously, the extended quantization group $\tilde{G}$ is a right principal bundle with structure group $U(1)$. Additional global features as well as constraints may be added by modifying the structure group, let me call it now $T$. For the sake of simplicity consider the case of the Heisenberg-Weyl group with one of the coordinates compactified. This would mean that the structure group $T$ will be chosen such that $\tilde{G}/T$ is a cylinder. The group law $g"=g'*g$ for $\tilde{G}$ is 
\begin{equation}
\begin{array}{c}
x"=x'+x\\
\\
p"=p'+p\\
\\
\zeta"=\zeta'\zeta e^{\frac{i}{h}[(1+\lambda)x'p+\lambda xp']}\\
\\
\end{array}
\end{equation}
where the first two transformations correspond to the non-extended Heisenberg-Weyl group $G$ while the last one is the group law associated to $U(1)$. We introduced another real parameter $\lambda$ to account for a complete class of central extensions differing in a coboundary generated by the function $\eta(x,p)=\lambda x p$. Here, the $T$-function condition generalizes the ordinary $U(1)$ equivariance in quantum mechanics, which is written as 
\begin{equation}
\Psi(\zeta*g)=\rho(\zeta)\Psi(g)
\end{equation}
where $\rho(\zeta)$ is the natural representation of $U(1)$ on the complex numbers, $\rho(\zeta)=\zeta$. If the group $T$ becomes larger we use a general representation $\mathcal{D}$ of $T=U(1)\cup T_{p}$ where $T_{p}$ is a maximal polarization subgroup of $T$. The T-function condition then becomes 
\begin{equation}
\Psi(g_{T}*g)=\mathcal{D}(g_{T})\Psi(g),\; \forall g_{T}\in T,\; \forall g\in \tilde{G}
\end{equation}
Following the example of reference [32] let $T=U(1)\times \{e_{k}, k\in \mathbb{Z}\}$ where $\{e_{k}, k\in \mathbb{Z}\}$ is the subgroup of $\tilde{G}$ of finite translations in the coordinate $x$ by an amount $kL$ where $L$ is a spatial period. Obviously such a fibration is trivial. Then we have $\mathcal{D}(\zeta,e_{k})=\zeta D(e_{k})$ where $D(e_{k})$ is a representation of $\{e_{k},k\in \mathbb{Z}\}$ in the complex numbers. There is an infinity of non-equivalent irreducible representations of the form $D^{\epsilon}(e_{k})=e^{\frac{i}{h}\epsilon k L}$ with $\epsilon \in [0,\frac{2\pi h}{L})$. There is a non-equivalent quantization associated with each choice of non-equivalent representation of $T$, parametrized by $\epsilon$. The $T$-function condition for the wavefunction then becomes
\begin{equation}
e^{\frac{i}{h}(1+\lambda)kLp} \Psi^{\epsilon}(x+kL,p,\zeta)=e^{\frac{i}{h}\epsilon k L}\Psi^{\epsilon}(x,p,\zeta)
\end{equation}
This set of non-equivalent quantizations can be reproduced by the introduction of an extra coboundary generated by the function $\epsilon x$, namely as a multiplicative factor of the form $e^{\frac{i}{h}\epsilon\frac{p'}{m}t}$ in the $\zeta\in U(1)$ composition law. At this moment topology change becomes relevant. When quantizing on a line, this term appears as a total derivative in the quantization $1$-form (or, equivalently in the Lagrangian) leading to actually equivalent quantum theories. However, when we change the topology and quantize the system on a circumference, the generating function $e^{\frac{i}{h}\epsilon x}$ is not single-valued on the circumference unless $\epsilon =\frac{2\pi h}{L}k,\;\forall k\in \mathbb{Z}$. Therefore two cocycles differing in a coboundary generated by $\epsilon x$ lead to non-equivalent theories on the circumference if $\epsilon \neq \frac{2\pi h}{L}k, k\in \mathbb{Z}$. This is a process of creation of non-trivial cohomology. 
It can be traced back to the appearance of two cocycles not physically equivalent despite being connected by a coboundary. It is therefore important to know what properties are topology-invariant and what properties change when the topology changes. In a way very similar to the language of differential forms and their coordinate independence, we can derive here a language relying on universal coefficient theorems, giving us a coefficient independent formulation that leads to those quantities conserved when a change in topology occurs.

First however, a prescription for performing a second quantization on a group is required. While I follow mainly ref. [34] there will be key additional points that will connect these ideas with the fact that non-trivial two-cocycles may be analyzed from the perspective of cohomology theories for which such cocycles may not be visible. This will lead us to the topology independent properties of the theory. The topology dependent part will be taken into account by means of extension groups in the universal coefficient theorem sequence. We need first a second quantization group $\tilde{G}^{(2)}$ representing the quantization group of the extended group $\tilde{G}$ defined for the construction of the polarized wavefunction previously. Given a Hilbert space $\mathcal{H}(\tilde{G})$ and its dual $\mathcal{H}^{*}(\tilde{G})$, one can define the direct sum 
\begin{equation}
\mathcal{F}(\tilde{G})=\mathcal{H}(\tilde{G})\oplus \mathcal{H}^{*}(\tilde{G})
\end{equation}
defined by the elements of the set 
\begin{equation}
\{\ket{f}=\ket{A}+\ket{B^{*}}; \ket{A}\in\mathcal{H}(\tilde{G}),\ket{B^{*}}\in \mathcal{H}^{*}(\tilde{G})\}
\end{equation}
Given group coordinates the elements above are defined such that the pairing $<\tilde{g}_{\mathcal{P}}^{*}|B^{*}>=<B|\tilde{g}_{\mathcal{P}}>=B^{*}_{\mathcal{P}}(\tilde{g})$ exists. The extended group $\tilde{G}$ acts on this vectorial space by means of the following rules ($\rho$ denotes the corresponding representation)
\begin{equation}
\rho(\tilde{g}')\ket{f}=\rho(\tilde{g}')\ket{A}+\rho(\tilde{g}')\ket{B^{*}}
\end{equation}
given that 
\begin{equation}
\bra{\tilde{g}_{\mathcal{P}}^{*}}\rho(\tilde{g}')\ket{B^{*}}=\bra{B}\rho^{\dagger}(\tilde{g}')\ket{\tilde{g}_{\mathcal{P}}}=B_{\mathcal{P}}^{*}(\tilde{g}'^{-1}*\tilde{g})
\end{equation}
The associated dual space is 
\begin{equation}
\mathcal{F}(\tilde{G})=\mathcal{H}^{*}(\tilde{G})\oplus \mathcal{H}^{**}(\tilde{G})
\end{equation}
with the associated set 
\begin{equation}
\{\bra{f}=\bra{A}+\bra{B^{*}}; \bra{A}\in\mathcal{H}^{*}(\tilde{G}),\bra{B^{*}}\in \mathcal{H}^{**}(\tilde{G})\cong \mathcal{H}(\tilde{G})\}
\end{equation}
with the adjoint action of $\tilde{G}$ defined in the usual way
\begin{equation}
\bra{f}\rho^{\dagger}(\tilde{g}')=\bra{A}\rho^{\dagger}(\tilde{g}')+\bra{B^{*}}\rho^{\dagger}(\tilde{g}')
\end{equation}
with 
\begin{equation}
\bra{B^{*}}\rho^{\dagger}(\tilde{g}')\ket{\tilde{g}^{*}_{\mathcal{P}}}=\bra{\tilde{g}_{\mathcal{P}}}\rho(\tilde{g}')\ket{B}
\end{equation}
The product of two arbitrary elements of $\mathcal{F}(\tilde{G})$ is 
\begin{equation}
<f'|f>=<A'|A>+<A'|B^{*}>+<B^{'*}|A>+<B^{'*}|B^{*}>=<A'|A>+<B^{'*}|B^{*}>
\end{equation}
because we know that 
\begin{equation}
\int_{\tilde{G}}\mu(\tilde{g})A^{'*}_{\mathcal{P}}(\tilde{g})B^{*}_{\mathcal{P}}(\tilde{g})=0=\int_{\tilde{G}}\mu(\tilde{g})B_{\mathcal{P}}'(\tilde{g})A_{\mathcal{P}}(\tilde{g})
\end{equation}
due to the integration over the central parameter $\zeta\in U(1)$. 
To the product space $\mathcal{M}(\tilde{G})=\mathcal{F}(\tilde{G})\otimes \mathcal{F}^{*}(\tilde{G})$ we can associate a symplectic structure which will make it a phase space
\begin{equation}
S(f',f)=\frac{-i}{2}(<f'|f>-<f|f'>)
\end{equation}
Such a phase space can be naturally embedded into a quantizing group 
\begin{equation}
\tilde{G}^{(2)}=\{\tilde{g}^{(2)}=(g^{(2)},\sigma)=(\tilde{g},\ket{f},\bra{f},\sigma)\}
\end{equation}
which is a central extension by $U(1)$ with parameter $\sigma$ of the semidirect product $G^{(2)}=\tilde{G}\otimes_{\rho}\mathcal{M}(\tilde{G})$ of the basic group $\tilde{G}$ and the phase space $\mathcal{M}(\tilde{G})$. The group law of $\tilde{G}^{(2)}$ is formally 
\begin{equation}
\begin{array}{c}
\tilde{g}"=\tilde{g}'*\tilde{g}\\
\\
\ket{f"}=\ket{f'}+\rho(\tilde{g}')\ket{f}\\
\\
\bra{f"}=\bra{f'}+\bra{f}\rho^{\dagger}(\tilde{g}')\\
\\
\sigma"=\sigma'\sigma e^{i\xi^{(2)}(g^{(2)'},g^{(2)})}\\
\\
\end{array}
\end{equation}
where the exponent is a cocycle defined as 
\begin{equation}
\xi^{(2)}(g^{(2)'},g^{(2)})=kS(f',\rho(\tilde{g}')f)
\end{equation}
where $k$ compensates for possible dimensions of $S$. The unitarity of $\rho$ guarantees the hermiticity of space-time symmetry operators. A choice of coordinates for the second quantization group $\tilde{G}^{2}$ corresponds to a choice of representation associated with a given polarization $\mathcal{P}$. 
\begin{equation}
\begin{array}{cc}
f_{\mathcal{P}}^{(+)}(\tilde{g})=<\tilde{g}_{\mathcal{P}}|f> & f_{\mathcal{P}}^{(-)}(\tilde{g})=<\tilde{g}^{*}_{\mathcal{P}}|f>\\
\\
f_{\mathcal{P}}^{*(+)}(\tilde{g})=<f|\tilde{g}^{*}_{\mathcal{P}}> & f_{\mathcal{P}}^{*(-)}(\tilde{g})=<f|\tilde{g}_{\mathcal{P}}>\\

\end{array}
\end{equation}
Such a splitting of $f$ represents the group analogue of the decomposition of fields in positive and negative frequency parts. By means of the closure relation 
\begin{equation}
1=\int_{\tilde{G}}\mu(\tilde{g})\{\ket{\tilde{g}_{\mathcal{P}}}\bra{\tilde{g}_{\mathcal{P}}}+\ket{\tilde{g}_{P}^{*}}\bra{\tilde{g}_{\mathcal{P}}^{*}}\}
\end{equation}
for $\mathcal{F}(\tilde{G})$, the cocycle in this coordinate system becomes (ignoring the semidirect action of $\tilde{G}$)
\begin{equation}
\begin{array}{c}
\xi^{(2)}(g^{(2)'},g^{(2)})=\frac{-ik}{2}\int\int_{\tilde{G}}\mu(\tilde{g}')\mu(\tilde{g})\{f^{'*(-)}(\tilde{g}')\Delta^{+}_{\mathcal{P}}(\tilde{g}',\tilde{g})f_{\mathcal{P}}^{+}(\tilde{g})-\\
\\
-f^{*(-)}_{\mathcal{P}}(\tilde{g}')\Delta_{\mathcal{P}}^{(+)}(\tilde{g}',\tilde{g})f^{'(+)}(\tilde{g})+f^{'*(+)}_{\mathcal{P}}(\tilde{g}')\Delta_{\mathcal{P}}^{-}(\tilde{g}',\tilde{g})f^{(-)}_{\mathcal{P}}(\tilde{g})-\\
\\
-f^{*(+)}(\tilde{g}')\Delta_{P}^{(-)}(\tilde{g}',\tilde{g})f^{'(-)}(\tilde{g})\}\\
\\
\end{array}
\end{equation}
where 
\begin{equation}
\begin{array}{c}
\Delta_{\mathcal{P}}^{(+)}(\tilde{g}',\tilde{g})=<\tilde{g}'|\tilde{g}_{\mathcal{P}}>=\sum_{n\in I}\psi_{\mathcal{P},n}(\tilde{g}')\psi^{*}_{\mathcal{P},n}(\tilde{g})\\
\\
\Delta_{\mathcal{P}}^{(-)}(\tilde{g}',\tilde{g})=<\tilde{g}^{'*}|\tilde{g}^{*}_{\mathcal{P}}>=\Delta_{\mathcal{P}}^{(+)}(\tilde{g},\tilde{g}')\\
\\
\end{array}
\end{equation}
The vector fields associated to the coordinates defined in the splitting are canonically conjugated
\begin{equation}
[\tilde{X}^{L}_{f^{*(-)}_{\mathcal{P}}(\tilde{g}')},\tilde{X}^{L}_{f_{\mathcal{P}}^{(+)}(\tilde{g})}]=k\Delta_{\mathcal{P}}^{(+)}(\tilde{g}',\tilde{g})\tilde{X}^{L}_{\sigma}
\end{equation}
\begin{equation}
[\tilde{X}^{L}_{f^{*(+)}_{\mathcal{P}}(\tilde{g}')},\tilde{X}^{L}_{f_{\mathcal{P}}^{(-)}(\tilde{g})}]=k\Delta_{\mathcal{P}}^{(-)}(\tilde{g}',\tilde{g})\tilde{X}^{L}_{\sigma}
\end{equation}
The functions $\Delta_{\mathcal{P}}^{(+)}(\tilde{g}',\tilde{g})$ and $\Delta_{\mathcal{P}}^{(-)}(\tilde{g}',\tilde{g})$ are the analogue of the propagators of quantum field theory. In the group theoretical approach these represent the central matrices of the cocycle. It is well known that the propagators in two different parametrizations of $\tilde{G}^{(2)}$ corresponding to two different polarization subalgebras $\mathcal{P}_{1}$ and $\mathcal{P}_{2}$ of $\tilde{\mathcal{G}}^{L}$ are related through polarization changing operators 
\begin{equation}
\begin{array}{c}
\Delta_{\mathcal{P}_{2}}^{(\pm)}(\tilde{h}',\tilde{h})=\int\int_{\tilde{G}}\mu(\tilde{g}')\mu(\tilde{g})\Delta^{\pm}_{\mathcal{P}_{2}\mathcal{P}_{1}}(\tilde{h}',\tilde{g}')\Delta^{\pm}_{\mathcal{P}_{1}}(\tilde{g}',\tilde{g})\Delta^{\pm}_{\mathcal{P}_{1}\mathcal{P}_{2}}(\tilde{g},\tilde{h})\\
\\
\Delta^{+}_{\mathcal{P}_{i}\mathcal{P}_{j}}(\tilde{h},\tilde{g})=\Delta_{\mathcal{P}_{i}\mathcal{P}_{j}}(\tilde{h},\tilde{g})\\
\\
\Delta^{-}_{\mathcal{P}_{i}\mathcal{P}_{j}}(\tilde{h},\tilde{g})=\Delta_{\mathcal{P}_{i}\mathcal{P}_{j}}(\tilde{g},\tilde{h})\\
\\
\end{array}
\end{equation}
The difference with respect to the standard quantum field theoretical method is that here, the integration is performed over the whole extended group $\tilde{G}$. The information about the propagation of the field is contained in the two co-cycle above and the fact that it implies the integration over the whole group $\tilde{G}$ and not just on a Cauchy surface $\Sigma\subset Q$ offers a higher level of generality. If however a reduction is employed $\int_{\tilde{G}}\mu(\tilde{g})\rightarrow \int_{\Sigma}d\sigma_{\nu}$, one may arrive at a loss of unitarity. Of course, the integration over the whole group $\tilde{G}$ must take into account its global properties and therefore the cohomology capable of detecting the cocycle inducing the propagator becomes relevant. The sensitivity to such a cocycle must carefully be adjusted by means of the chosen coefficients. In the case when a horizon forms, the required covariance will induce special choices of coefficients which will trigger additional global factors in the pairing defining the inner product used in the Hawking calculation. 


Using the Fourier-like parametrization associated with the basis $\mathcal{B}(\mathcal{F}(\tilde{G}))=\{\ket{n}+\ket{m^{*}}\}$ made of Hamiltonian eigenstates, if we denote the Fourier coefficients of particles and anti-particles as
\begin{equation}
\begin{array}{cc}
a_{n}=<n|f>, & b_{m}=<m^{*}|f>\\
a^{*}_{n}=<f|n>, & b_{m}^{*}=<f|m^{*}>\\
\end{array}
\end{equation}
and we write a polarization subalgebra 
\begin{equation}
\mathcal{P}^{(2)}=<\tilde{X}^{L}_{a_{n}}, \tilde{X}^{L}_{b_{m}}; \mathcal{G}_{\Theta^{(2)}}\cong \tilde{G}^{L}>
\end{equation}
we obtain the basic operators of the theory: annihilation operators of particles and antiparticles $\hat{a}_{n}=\tilde{X}^{R}_{a_{n}^{*}}$, $\hat{b}_{m}=\tilde{X}^{R}_{b_{m}^{*}}$, and the corresponding creation operators 
$\hat{a}_{n}^{\dagger}=-\frac{1}{k}\tilde{X}^{R}_{a_{n}}$, $\hat{b}_{m}^{\dagger}=-\frac{1}{k}\tilde{X}^{R}_{b_{m}}$.


The two fields appearing in the inner product
\begin{equation}
(\phi_{1},\phi_{2})=i\int d^{n-1}x|g|^{1/2}g^{0\nu}\varphi_{1}^{*}(x,t)\overleftrightarrow{\partial_{\nu}}\varphi_{2}(x,t)
\end{equation}
are solutions of the wave equation. Their form is correct and needs no modification when a topology change occurs. However, the general pairing of such fields in the context of a non-trivial topology must take into account the universal coefficient theorem. 

One may ask what happens if a structure of this form is used in order to map a space before and after the collapse of a dust cloud into a black hole. If the definition of the topology is such that points separated by a horizon are not defined to belong in the same open set then modifications must be implemented. 
Let $G$ and $K$ be two abstract groups. A group $\tilde{G}$ is said to be an extension of $G$ by $K$ if $K$ is an invariant subgroup of $\tilde{G}$ and $\tilde{G}/K=G$.
In terms of exact sequences this means that 
\begin{equation}
1 \rightarrow K \rightarrow \tilde{G}\rightarrow G \rightarrow 1
\end{equation}
is exact i.e. $K$ is injected into $\tilde{G}$ and $\tilde{G}$ is projected onto $G$ by the canonical homomorphism so that $G=\tilde{G}/K$. However, the mere knowledge of $K$ and $G$ does not define $\tilde{G}$ uniquely. In order to be able to discern extensions one has to define two exact sequences

\begin{equation}
1\rightarrow K \xrightarrow{i_{1}}\tilde{G}_{1}\xrightarrow{\pi_{1}} G \rightarrow 1
\end{equation}
\begin{equation}
1\rightarrow K \xrightarrow{i_{2}}\tilde{G}_{2}\xrightarrow{\pi_{2}} G \rightarrow 1
\end{equation}
If the two group extensions are related via an isomorphism $\tilde{f}$: 
\begin{equation}
\tilde{f}:\tilde{G}_{1}\rightarrow \tilde{G}_{2}
\end{equation}

and the injective maps $i_{1,2}$ and the projections $\pi_{1,2}$ satisfy
\begin{equation}
\begin{array}{c}
i_{2}=\tilde{f}\circ i_{1}\\
\\
\pi_{1}=\pi_{2}\circ \tilde{f}\\
\\
\end{array}
\end{equation}
then the extensions are equivalent. 
Consider now the two group extensions, defined by two different two-cocycles $\xi_{1}$ and $\xi_{2}$ with their group laws defined separately with simple brackets (...) for the first group and square brackets [...] for the second group: 
\begin{widetext}
\begin{equation}
\begin{array}{cc}
(g', \theta')(g, \theta)=(g'g, \theta'+\theta+\xi_{1}(g',g)), \; & [g', \theta'][g, \theta]=[g'g, \theta'+\theta+\xi_{2}(g',g)]
\end{array}
\end{equation}
\end{widetext}
If there exists an isomorphism $\tilde{f}$ as defined above and if we can rewrite 
\begin{equation}
(g, \theta)=(e, \theta)(g, 0)
\end{equation}
$(e, 0)$ being the identity of this law, $\tilde{f}$ is completely determined when the images of $(e, \theta)$ and $(g, 0)$ are given. 
From the conditions on the injection and projection above one obtains 
\begin{equation}
\begin{array}{c}
\tilde{f}\circ i_{1}=i_{2} \; \Rightarrow \; \tilde{f}(e, \theta)=[e, \theta]\\
\\
\pi_{2}\circ \tilde{f}=\pi_{1} \; \Rightarrow \; \tilde{f}(g, 0)=[g, \eta(g)]\\
\\
\end{array}
\end{equation}
This implies a general form for $\tilde{f}$ namely 
\begin{equation}
\tilde{f}(g, \theta)=[g, \theta+\eta(g)]
\end{equation}
The knowledge of $\eta$ determines the knowledge of $\tilde{f}$. However, $\tilde{f}$ is also a homomorphism hence
\begin{widetext}
\begin{equation}
\tilde{f}(g'g, \theta' + \theta + \xi_{1}(g',g))=[g'g, \theta' + \theta + \xi_{1}(g',g)+\eta(g'g)]
\end{equation}
\end{widetext}
must be equal to 
\begin{equation}
\begin{array}{c}
\tilde{f}(g', \theta')\tilde{f}(g, \theta)=[g', \theta'+\eta(g')][g, \theta+\eta(g)]=\\
\\
=[g'g, \theta'+\theta+\xi_{2}(g',g)+\eta(g')+\eta(g)]\\
\\
\end{array}
\end{equation}
and hence
\begin{equation}
\begin{array}{c}
\xi_{1}(g',g)=\xi_{2}(g',g)+\eta(g')+\eta(g)-\eta(g'g)=\\
\\
=\xi_{2}(g',g)+\xi_{cob}(g',g)\\
\\
\end{array}
\end{equation}
where the notation $\xi_{cob}(g',g)$ is used for the two-coboundary generated by $\eta(g)$.
The calculation above gives a condition for the equivalence of extensions. One can see that proportional two-cocycles 
$\xi_{2}=\lambda \xi_{1}$ may define equivalent groups but inequivalent extensions. In the case of the black hole formation the inequivalent extensions are those considered when constructing the pairings resulting in the definition of the out-going radiation.

Therefore what we require is to have on one side the trivial situation far away from a horizon, and, on the other side of the pairing, the non-trivial situation, close to the horizon. We do not demand triviality or non-triviality of the cohomology in any of these cases. What is required however is that no matter what the sensitivity of cohomology, the pairing between homology and cohomology must remain covariant and objectively take into account such a transition from a region characterized by one topology to another region characterized by another. In this case we obtain a non-trivial factor that will correct the thermal nature of the out-coming radiation. 
In order to make the connection with the bracket construction and to classify the extensions one has to rely on a fiber bundle definition of the extension. Let therefore $G$ and $K$ be abstract general groups and $\tilde{G}$ be the extension of $G$ by $K$. One can relate the cosets of $K$ in $\tilde{G}$, each defining an element $g\in G$ with the fibers over $g$ of a fiber bundle that defines the extension. The fiber through $\tilde{g}_{0} \in \tilde{G}$ is given by 
\begin{equation}
\pi^{-1}(\pi(\tilde{g}_{0}))=\{\tilde{g}|\tilde{g}=k\tilde{g}_{0},k\in K\}
\end{equation}
A section of $\tilde{G}(K,\tilde{G}/K=G)$
\begin{equation}
\begin{array}{cc}
s:G\rightarrow \tilde{G}, \; & s: (g)\rightarrow s(g)\\
\end{array}
\end{equation}
selects an element in $\tilde{G}$ in each fiber. 
Now, given a fiber
\begin{equation}
\pi(s(g''))=\pi(s(g')s(g))
\end{equation}
thus there exists a factor $\omega(g',g) \, \in K$ such that 
\begin{equation}
s(g')s(g)=\omega(g',g)s(g',g)
\end{equation}
and this relation defines the factor $\omega(g',g)$. 
One can define $\omega(g',e)=\omega(e,g)=s(e)$ and take $s(e)=\tilde{e} \in \tilde{G}$.
Thus, one obtains the normalized section. 
Similarly one can obtain, for a normalized section, also a normalized factor:
\begin{equation}
\omega(g,e)=\omega(e,g)=\omega(e,e)=e \in K
\end{equation}
As a general statement, relative to any normalized trivializing section $s: G \rightarrow \tilde{G}$ one can associate a factor system $\omega : G \times G \rightarrow K$ satisfying 
\begin{equation}
\omega(g'',g)\omega(g''g',g)=([s(g'')]\omega(g',g))\omega(g'',g'g)
\end{equation}
where $[s(g)]k=s(g)ks(g)^{-1}$ $\forall k \in K$.
According to this fiber bundle representation of the extensions, the group law of the group extension can be defined in terms of the factor system as 
\begin{equation}
(g'', k'')=(g', k')*_{s}(g,k)=(g'g, k'[s(g')]k\omega(g',g))
\end{equation}
Returning to the physical problem, the invariant bracket defined above,
\begin{equation}
(\phi_{1},\phi_{2})=i\int d^{n-1}x|g|^{1/2}g^{0\nu}\phi_{1}^{*}(x,t)\overleftrightarrow{\partial_{\nu}}\phi_{2}(x,t)
\end{equation}
 must be extended in order to obtain a topologically covariant description. The definition of the adjoint of the topological bracket can be identified as the right hand side of the universal coefficient theorem. When a choice of coefficients is considered such that the horizon of the black hole becomes visible one obtains a correction to the bracket as given by the factor that characterizes the extension of the homology group in a dimension smaller by one unit. It will be this extension that will generate the quantization prescription to be used in the physical situation. The bracket is defined now with a correction in the group operation associated to its defining symmetry. Hence a topological factor is missing in the construction used in [1]. I underline that this factor is purely topological.
Hence one has to extend the scalar bracket when a topological covariance is required:
\begin{equation}
(\phi_{1},\phi_{2})'=<\phi_{1},\phi_{2}>\omega(\phi_{1},\phi_{2})(\phi_{1},\phi_{2})
\end{equation}
where the $<...>$ notation refers to the topological invariant and $\omega(\phi_{1},\phi_{2})$ refers to the factor system that characterizes the extension and depends on the choice of the coefficient structure. 
This factor will appear also in the coefficients defining the probability of particle detection far from the black hole horizon. 
I must add that this is the first derivation of the fact that such a transformation is required, using almost only topological arguments. This, by itself is a very important conclusion. However, I am aware that a detailed calculation of fluxes might also be beneficial. This will be the subject of a future work. 

To make these considerations more accurate I will follow [1]. These results can also be interpreted in terms of the group approach to the second quantization. Consider the vacuum state at the infinite past as 
\begin{equation}
|0_{-}>=\sum\sum\lambda_{AB}|A_{I}>|B_{H}>
\end{equation}
where $|A_{I}>$ is the outgoing state with $n_{ja}$ particles in the $j$th outgoing mode and $|B_{H}>$ is the horizon state with $n_{kb}$ particles in the $k$th mode going into the hole. 
Otherwise stated 
\begin{equation}
\begin{array}{c}
|A_{I}> = \prod_{j}(n_{ja}!)^{-1/2}(b^{+}_{j})^{n_{ja}}|0_{I}>\\
\\
|B_{H}> = \prod_{k}(n_{kb}!)^{-1/2}(c^{+}_{k})^{n_{kb}}|0_{I}>\\
\\
\end{array}
\end{equation}
One can chose an observable at the far future, composed only of $\{b_{j}\}$ and $\{b_{j}^{+}\}$ and operating only on the vectors $|A_{I}>$. 
The expectation value of this observable can be written as 
\begin{equation}
<0_{-}|Q|0_{-}>=\sum\sum\rho_{AC}Q_{CA}
\end{equation}
where $Q_{CA}=<C_{I}|Q|A_{I}>$ is the matrix element of the observable in the Hilbert space of the outgoing states. 
The density matrix is 
\begin{equation}
\rho_{AC}=\sum\lambda_{AB}\bar{\lambda}_{CB}
\end{equation}
and is associated to measurements in the far future but not to measurements of systems falling into the black hole. 
But, as has been shown above, the propagators for a transition from incoming matter to out-going radiation 
\begin{equation}
\begin{array}{c}
\Delta_{\mathcal{P}}^{(+)}(\tilde{g}',\tilde{g})=<\tilde{g}'|\tilde{g}_{\mathcal{P}}>=\sum_{n\in I}\psi_{\mathcal{P},n}(\tilde{g}')\psi^{*}_{\mathcal{P},n}(\tilde{g})\\
\\
\Delta_{\mathcal{P}}^{(-)}(\tilde{g}',\tilde{g})=<\tilde{g}^{'*}|\tilde{g}^{*}_{\mathcal{P}}>=\Delta_{\mathcal{P}}^{(+)}(\tilde{g},\tilde{g}')\\
\\
\end{array}
\end{equation}
are matrices associated to non-trivial two-cocycles detected by cohomology with certain coefficients. Writing therefore the associated matrix elements in a topology covariant way implies adding additional factors correcting precisely the distribution of the outgoing radiation. But the visibility of the cocycles is controlled by the coefficients in cohomology.

It is at this point where several extensions of the standard prescription are necessary. 
The above density matrix does not encode the full information that can be obtained in the far future. It does encode however everything that can be obtained from non-topological considerations.


A particular form of the universal coefficient theorem is 
\begin{widetext}
\begin{equation}
0\rightarrow Ext(H_{i-1}(G;R),M)\rightarrow H^{i}(G;M)\xrightarrow{h} Hom(H_{i}(G;R),M)\rightarrow 0
\end{equation}
\end{widetext}
This can be interpreted in a form that resembles the interpretation of the non-commutativity of some physical observables: 
the third arrow 
\begin{equation}
H^{i}(G;M)\xrightarrow{h} Hom(H_{i}(G;R),M)
\end{equation}
maps the cohomology with coefficients in the group $M$ into the homomorphisms between the homology with coefficients in $R$ and the group $M$. The sequence is exact, hence this map is a surjection. This means there are no elements in the set of homomorphisms from the homology with coefficients in $R$ to the group $M$ not represented in the cohomology with coefficients in $M$. However, there are elements in the cohomology that can be mapped into the same element of $Hom$. The second arrow
\begin{equation}
Ext(H_{i-1}(G;R),M)\rightarrow H^{i}(G;M)
\end{equation}
is an injection. Hence the extension encodes the way in which the use of a coefficient structure instead of another changes the classes of the cohomology.
This implies a change in the factors of the inner products used. 

One may ask if locality is preserved in this situation. Indeed, the problem of locality when unitarity is restored appears to be fundamental to the AdS/CFT solution of the information paradox [6], [8].  The information, in the approach of this work, is encoded in the global topological structure of the field in such a way that it is not accessible by any local measurements. One has to remember that the quantum field is not a measurable quantity. There is no physically observable "quantum field" in the same way in which there is no physically observable wavefunction. Nevertheless, the global, topological properties of the fields (and wavefunctions) are important and encode relevant information. Any local measurement can be seen as a "small" (weak) measurement. Can such a measurement reveal the global information? The correct answer to this question is no. Any weak measurement will reveal a weak information that will not provide any access to the information encoded globally and retrievable only via a statistical topological measurement. If one choses a coefficient structure for which the global non-triviality is invisible, locality is regained. Information is conserved but only in the factors appearing due to the use of the extension group. Hence unitarity is still preserved but in a "hidden" form (in the extension). If one choses a suitable coefficient structure the global information becomes accessible due to the manifest visibility of the global non-triviality. However, one cannot recover the information unless one performs a probing of the topology. This may look non-local in a sense but the information obtained in this way concerns topologically non-trivial field (wavefunction) structures hence this "non-locality" is not a physical one but rather one related to a choice of performing certain measurements. 


\section{5. conclusion}
As a conclusion, I have shown that topological corrections to the thermal radiation of a black hole as given by the requirement of topological covariance of the laws of physics can account for a factor in the coefficients defining the thermal radiation. This factor imposes non-trivial changes in the form of the distribution function that amount to non-thermal corrections. 
This observation confirms the suspicions that the solution of the unitarity problem relies on non-perturbative effects and on topological properties of the quantum groups involved in the derivation of the radiation distribution function. 
\\
\section{bibliography}


\begin{thebibliography}{9}
\bibitem{1}
S. Hawking, Phys. Rev. D 10 (14), 2460 (1976)
\bibitem{2}
S. Hawking, Commun. Math. Phys. 43, 199 (1975)
\bibitem{3}
A. Einstein, Ann. d. Physik 354 (7), 769-822 (1916)
\bibitem{4}
J. A. Azcarraga, M. Josi, Lie groups, Lie algebras, cohomologies and some applications in physics, 
\par See page 291 for the connection between the topological structure of the Galilei and Poincare groups and the existence of a simple covariant formulation. 
\bibitem{5}
 Schwarzschild, K.  Sitzungsberichte der K\"oniglich Preussischen Akademie der Wissenschaften 7: 189Ð196 (1916)
 \bibitem{6}
D. A. Lowe, J. Polchinsky, L. Thorlacius, J. Uglum, Phys. Rev. D 52 6997 (1995)
\bibitem{7}
N. Bogoliubov, J. Phys. (USSR), 11, p. 23 (1947)
\bibitem{8}
J. M. Maldacena, Adv. Theor. Math. Phys. 2, 231
\bibitem{9} 
A. T. Patrascu, Phys. Rev. D 90, 045018 (2014)
\bibitem{10}
J. F. Davis, P. Kirk, Lecture Notes in Algebraic Topology (see page 43 and 47 in notes)
\bibitem{11}
S. Willard, General Topology, Dover Books, ISBN 9780486434797 (1970)
\bibitem{12}
J. L. Kelley, General Topology, Springer Graduate Texts in Mathematics, ISBN 9780387901251 (1975)
\bibitem{13}
E. C. Zeeman, J. Math. Phys. 5, 490 (1964)
\bibitem{14}
E. C. Zeeman, Topology 6. 161 (1967)
\bibitem{15}
S. W. Hawking, A. R. King, P. J. McCarthy, J. Math. Phys. 17, 174 (1976)
\bibitem{16}
R. G\"obel, Physikal Teil (II) der Habilitationsschrift, Wurzburg (1973)
\bibitem{17}
For example: T. Leinster, Course in General topology, pag. 13 (2014)
\bibitem{18}
A. Hatcher, Algebraic Topology, Cambridge University Press, ISBN 052179160-X (2002)
\bibitem{19}
C. R. F. Maunder, Algebraic Topology, Cambridge University Press, ISBN 0486691314 (1996)
\bibitem{20}
H. Sato, Algebraic Topology: An Intuitive Approach, American Mathematical Society, ISBN 0821810464 (1999)
\bibitem{21}
L. A. Steen, J. A. Seebach, Counterexamples in Topology, Dover Publications, ISBN 048668735X (1978)
\bibitem{22}
G. T. Horowitz, Class. Quantum Grav. 8 587 (1991)
\bibitem{23}
S. B. Giddings, A. Strominger, Nuclear Physics B, Vol. 306, Issue 4, Pag. 890Ð907 (1988)
\bibitem{24}
B. Harms, Y. Leblanc, Annals of Physics, Vol 242, Issue 2, Pag. 265Ð274 (1995)
\bibitem{25} 
J. L. F. Barbon, Journal of Physics: Conference Series 171,  012009 (2009)
\bibitem{26}
R. Kirby, The Topology of 4-manifolds. Springer Lecture Notes 1374 (1989)
\bibitem{27}
M. H. Freedman, J. Differential Geom. 17,  no. 3, pag. 357Ð453 (1982)
\bibitem{28}
C. Manolescu, arXiv/math : 1303.2354
\bibitem{29}
J. Guerrero, J. L. Jaramillo, V. Aldaya, J. Math. Phys. Vol. 45, 2051 (2004)
\bibitem{30} 
V. Aldaya, J. Guerrero, G. Marmo, Quantization on a Lie group: Higher order polarizations, Proc. Intern. Sympos. of Symm. in Science X, ISBN 978-1-1539-9 (1998)
\bibitem{31}
O. Viro, J. of Knot Theory and its Ramifications, Vol. 18, No. 6, pag. 729 (2009)
\bibitem{32} 
V. Aldaya, M. Calixto, J. Guerrero, Commun. Math. Phys. Vol. 178, pag. 399 (1999)
\bibitem{33} 
V. Aldaya, M. Calixto, J. M. Cervero, Commun. Math. Phys. 200, pag. 325 (1999)
\bibitem{34}
M. Calixto, V. Aldaya, M. Navarro, Int. J. of Mod. Phys. A, Vol. 15, 25 (2000)
\bibitem{35}
S. A. Fulling, Phys. Rev. D7, 2850 (1973)
\bibitem{36}
J. L. Loday, J. of Algebra, 181, No. 0127, pag. 414 (1996)
\bibitem{37}
E. J. Saletan, J. Math. Phys. 2, 1 (1961)
\end{thebibliography}
\end{document}